\documentclass[intlimits,twoside,a4paper]{article}
\usepackage{amsmath,amssymb}
\usepackage{graphicx}
\usepackage{wrapfig}
\usepackage[T2A]{fontenc}
\usepackage[cp1251]{inputenc}

\usepackage{threeparttable}

\usepackage{cmpj}

\issue{2011}{14}{4}{43701}

\doinumber{10.5488/CMP.14.43701}

\title[Theoretical studies of $^{63}\text{Cu}$  Knight shifts of the normal state
of  $\text{YBa}_{2}\text{Cu}_{3}\text{O}_{7}$]{Theoretical studies of $^{63}\text{Cu}$  Knight shifts of the normal state
of  $\text{YBa}_{2}\text{Cu}_{3}\text{O}_{7}$}

\author[M.Q. Kuang \textsl{et al.}]{M.Q. Kuang\refaddr{ad1}\thanks{E-mail: mqkuang@yeah.net}\, , S.Y.~Wu\refaddr{ad1,ad2}, Z.H.~Zhang\refaddr{ad1}, B.T.~Song\refaddr{ad1}}

\authorcopyright{M.Q. Kuang, S.Y.~Wu, Z.H.~Zhang, B.T.~Song, 2011}

\addresses{\addr{ad1}Department of Applied Physics, University of
Electronic Science and Technology of China, \\ 610054~Chengdu,
P.R. China%
\addr{ad2} International Centre for Materials Physics,
Chinese Academy of Science,  110016~Shenyang, P.R. China}

\date{Received June 7, 2011, in final form August 15, 2011}

\begin{document}
\maketitle
\begin{abstract}

The $^{63}\text{Cu}$ Knight shifts and $g$ factors for the
normal state of $\text{YBa}_{2}\text{Cu}_{3}\text{O}_{7}$ in
tetragonal phase are theoretically studied in a uniform way from the
high (fourth-) order perturbation formulas of these parameters for a
$3d^{9}$ ion under tetragonally elongated octahedra. The
calculations are quantitatively correlated with the local structure
of the $\text{Cu}^{2+}(2)$ site in
$\text{YBa}_{2}\text{Cu}_{3}\text{O}_{7}$. The theoretical results
show good agreement with the observed values, and the improvements
are achieved by adopting fewer adjustable parameters as compared to
the previous works. It is found that the significant anisotropy of
the Knight shifts is mainly attributed to the anisotropy of the $g$
factors due to the orbital interactions.


\pacs 71.70.Ch, 74.25.Nf, 74.72.Bk
\end{abstract}

\section{Introduction}

$\text{YBa}_{2}\text{Cu}_{3}\text{O}_{7}$ has become an
important topic in the microwave~\cite{1}, magnetism~\cite{2,3} as
well as nuclear quadrupole resonance (NQR) and nuclear magnetic
resonance (NMR)~\cite{4,5} researches. In particular, the Knight
shifts as the results of NMR experiments can yield important
information about the local environments and electronic distribution
around the magnetic nuclei. For example, the Knight shifts
$K_{\parallel}$ and $K_{\perp}$ as well as the anisotropic $g$
factors $g_{\parallel}$ and $g_{\perp}$ were measured~\cite{6,7} for
the tetragonal $\text{Cu}^{2+}(2)$ site in
 $\text{YBa}_{2}\text{Cu}_{3}\text{O}_{7}$ of normal state at room temperature.
However, the theoretical studies~\cite{6} on the Knight shifts seem
unsatisfactory. Firstly, the previous calculations of the Knight
shifts were generally based on the simple second-order perturbation
formulas, while the contributions from the higher (third and fourth)
order perturbation terms were not taken into account. Secondly, the
analysis on the Knight shifts was not correlated to the local
structure of the magnetic site, but was treated by introducing various
adjustable parameters (e.g., the related energy separations).
Finally, the $g$ factors were not quantitatively treated in a
uniform way. In order to study the Knight shifts and the $g$ factors
of
 $\text{YBa}_{2}\text{Cu}_{3}\text{O}_{7}$ to a better extent, the high (fourth-) order
perturbation formulas of the Knight shifts and the $g$ factors for a
$3d^{9}$ ion located in tetragonally elongated octahedra are
applied to the tetragonal $\text{Cu}^{2+}(2)$ site of
 $\text{YBa}_{2}\text{Cu}_{3}\text{O}_{7}$ in this work, and the local structure of
this site is quantitatively involved in the calculations.

\section{Calculation}

The anisotropic Knight shifts and $g$ factors may be ascribed
to the tetragonal $\text{Cu}^{2+}(2)$ site in
$\text{YBa}_{2}\text{Cu}_{3}\text{O}_{7}$, which is coordinated to
five oxygen ions forming a tetragonally elongated octahedron
(i.e., one of the apical ligands in the oxygen octahedron is moved
to infinity)~\cite{8} As regards a $\text{Cu}^{2+}(3d^{9})$ ion in
tetragonally elongated octahedra, its original cubic $^{2}E_{\mathrm{g}}$
ground state would be separated into two orbital singlets
$^{2}B_{1{\mathrm{g}}}$ (or $|x^{2}-y^{2}\rangle$) and $^{2}A_{1{\mathrm{g}}}$ (or $|z^{2}\rangle$),
and the former is the lowest level. Meanwhile, the original cubic
$^{2}T_{2{\mathrm{g}}}$ exited state may spilt into an orbital singlet
$^{2}B_{2{\mathrm{g}}}$ (or $|xy\rangle$) and a doublet $^{2}E_{\mathrm{g}}$ (or $|yz\rangle$ and
$|xz\rangle$)~\cite{9}. In order to make uniform investigations of the
Knight shifts and the $g$ factors as well as to overcome the shortcomings
of the previous study~\cite{6} based on the simple second-order
perturbation formulas, the high order perturbation $g$ formulas
\cite{10} including the third- and fourth- order perturbation terms
for a tetragonally elongated octahedral $3d^{9}$ cluster are adopted
here. Applying the Macfarlane's perturbation loop method~\cite{11},
the perturbation Hamiltonian for an orthorhombic $3d^{9}$ cluster
under external magnetic field can be written as follows:
\begin{eqnarray}
H'&=&H_{\mathrm{SO}}(\zeta ) + H_{\mathrm{Ze}}(\emph{k}) + H_{\mathrm{hf}}(\emph{P})   ,
\label{eq1}
\end{eqnarray}
where $H_{\mathrm{SO}}$, $H_{\mathrm{Ze}}$ and $H_{\mathrm{hf}}$ are, respectively, the
spin-orbit coupling interactions, the Zeeman term and the hyperfine
interactions, with the corresponding spin-orbit coupling coefficient
$\zeta$, the orbital reduction factor \emph{k} and the dipolar
hyperfine structure parameter \emph{P} for a $3d^{9}$ ion in
crystals. Utilizing the perturbation method~\cite{11}, the
perturbation formulas for an orthorhombically elongated $3d^{9}$
cluster can be derived as follows~\cite{10}:
\begin{eqnarray} g_{\parallel}&=&g_{\mathrm{s}}+8\emph{k}\frac{\zeta}{E_{1}}+\emph{k}\frac{\zeta^{2}}{E^{2}_{2}}
+4\emph{k}\frac{\zeta^{2}}{E_{1}E_{2}}-g_{\mathrm{s}}\zeta^{2}\left( \frac{1}{E^{2}_{1}}-\frac{1}{2E^{2}_{2}}\right)\nonumber\\ %
&+&\emph{k}\zeta^{3}
{\left(\frac{4}{E_{1}}-\frac{1}{E_{2}}\right)}\frac{1}{E^{2}_{2}}
-2\emph{k}\zeta^{3} \left(\frac{2}{E^{2}_{1}E_{2}}-\frac{1}{E_{1}E^{2}_{2}}\right)+g_{\mathrm{s}}\zeta^{3} \left(\frac{1}{E_{1}E^{2}_{2}}-\frac{1}{2E^{3}_{2}}\right),\nonumber\\[2ex] %
 g_{\perp}&=&g_{\mathrm{s}}+2\emph{k}\frac{\zeta}{E_{2}}-4\emph{k}\frac{\zeta^{2}}{E_{1}E_{2}}+ \emph{k}\zeta^{2}{\left(\frac{2}{E_{1}}-\frac{1}{E_{2}}\right)}\frac{1}{E_2}+2g_{\mathrm{s}}\frac{\zeta^{2}}{E^{2}_{1}}\nonumber\\
&+&\emph{k}\zeta^{3}
{\left(\frac{2}{E_{1}}-\frac{1}{E_{2}}\right)\left(\frac{1}{E_{2}}+\frac{2}{E_{1}}\right)}\frac{1}{2E_2}
- g_{\mathrm{s}}\zeta^{3}\left(\frac{1}{2E^{2}_{1}E_{2}}-\frac{1}{2E_{1}E^{2}_{2}}+\frac{1}{2E^{3}_{2}}\right),
\label{eq2}\end{eqnarray}
here $g_{\mathrm{s}}(\approx 2.0023)$ is the pure spin value.
\emph{k} is the orbital reduction factor. $\zeta$ is the spin-orbit
coupling coefficient for the $3d^{9}$ ion in crystals, which can be
expressed in terms of the corresponding free-ion value $\zeta_{0}$
as $\zeta\approx\emph{k} \zeta_{0}$. It is noted that when the
third- and fourth -order perturbation terms are neglected, i.e.,
only the first and second terms in the right sides of equation~\eqref{eq1} are
reserved, the above formulas may be reduced to those of the previous
work~\cite{6}. $E_{1}$ and $E_{2}$ are the energy separations
between the excited $^{2}B_{2{\mathrm{g}}}$ and $^{2}E_{\mathrm{g}}$ and the ground
$^{2}B_{1{\mathrm{g}}}$ states~\cite{12}:
\begin{eqnarray}
E_{1}&=&10D_{\mathrm{q}}\,,\nonumber\\
E_{2}&=&10D_{\mathrm{q}}-3D_{\mathrm{s}}+5D_{\mathrm{t}}\,.
\label{eq3}
\end{eqnarray}
Here $D_{\mathrm{q}}$ is the cubic field parameter, and $D_{\mathrm{s}}$ and
$D_{\mathrm{t}}$ are the tetragonal field parameters. The local structure of
the five-fold coordinated $\text{Cu}^{2+}(2)$ site in
$\text{YBa}_{2}\text{Cu}_{3}\text{O}_{7}$ can be described as the
parallel $\text{Cu}^{2+}-\text{O}^{2-}$ distance
$\emph{R}_{\parallel}$ ($\approx 2.426$ {\AA}) along $\emph{c}$ axis
and the four perpendicular distances $\emph{R}_{\perp}$($\approx
1.939$ {\AA}) along $\emph{a}$ and $\emph{b}$ axes~\cite{8}. Thus, the
tetragonal field parameters can be determined from the superposition
model~\cite{12}:
\begin{eqnarray}
D_{\mathrm{s}}&=&2 \overline{A}_{2}\emph{R}_{0} \frac{1}{7}{\left[\left(\frac{\emph{R}_{0}}{\emph{R}_{\parallel}}\right)^{t2}-2\left(\frac{\emph{R}_{0}}{\emph{R}_{\perp}}\right)^{t2}\right]}, \qquad
D_{\mathrm{t}}=16
\overline{A}_{4}\emph{R}_{0}
\frac{1}{21}{\left[\left(\frac{\emph{R}_{0}}{\emph{R}_{\parallel}}\right)^{t4}-2\left(\frac{\emph{R}_{0}}{\emph{R}_{\perp}}\right)^{t4}\right]}\,.
\label{eq4}\end{eqnarray}
Here $t_{2}\approx3$ and $t_{4}\approx5$ are the power-law
exponents~\cite{12}. $\overline{A}_{2}(\emph{R}_{0})$ and
$\overline{A}_{4}(\emph{R}_{0})$ are the intrinsic parameters, with
the reference bond length $\emph{R}_{0}$ taken as the average
$\text{Cu}^{2+}-\text{O}^{2-}$ distance, i.e., $\emph{R}_{0}=
\overline{R}=(\emph{R}_{\parallel}+4\emph{R}_{\perp})/5$. For
octahedral $3d^{n}$ clusters, the relationships
$\overline{A}_{4}(\emph{R}_{0}) \approx(3/4)D_{\mathrm{q}}$ and
$\overline{A}_{2}(\emph{R}_{0})\approx10.8\overline{A}_{4}(\emph{R}_{0})$
\cite{12} are proved valid for many systems and reasonably applied
here.

According to the optical spectra for $\text{Cu}^{2+}$ in
oxides~\cite{13}, the cubic field parameter $D_{\mathrm{q}}\approx1260$~cm$^{-1}$ and the orbital reduction factor $\emph{k}\approx0.76$ can
be obtained for the studied system by fitting the observed d-d
transitions. The spin-orbit coupling
coefficient is determined using the free-ion value $\zeta_{0} (\approx 829~$cm$^{-1}$~\cite{14}) for $\text{Cu}^{2+}$. Substituting these values into equation~\eqref{eq1},
the theoretical $g$ factors are calculated and shown in table~1.

Now we turn to the investigations of the Knight shifts. From
the relationships between the $g$ factors and the Knight shifts and
the perturbation procedure similar to that in Ref.~\cite{6}, the
high order perturbation formulas (i.e., those containing the third- and
fourth- order perturbation contributions) of the Knight shifts for
the tetragonal $\text{Cu}^{2+}(2)$ site in
$\text{YBa}_{2}\text{Cu}_{3}\text{O}_{7}$ can be expressed as:
\begin{eqnarray}
\emph{K}_{\parallel}&=&2\chi\biggl[\frac{8\emph{k}}{E_{1}}+\frac{\emph{k}\zeta}{E^{2}_{2}}+4\emph{k}\frac{\zeta}{E_{1}E_{2}}-
g_{\mathrm{s}}\zeta\left(\frac{1}{E^{2}_{1}}-\frac{1}{2E^{2}_{2}}\right)+\emph{k}\zeta^{2}
{\left(\frac{4}{E_{1}}-\frac{1}{E_{2}}\right)}\frac{1}{E^{2}_{2}}
\nonumber\\
&-&2\emph{k}\zeta^{2}\left(\frac{2}{E^{2}_{1}E_{2}}-\frac{1}{E_{1}E^{2}_{2}}\right)+g_{\mathrm{s}}\zeta^{2}
\left(\frac{1}{E_{1}E^{2}_{2}}-\frac{1}{2E^{3}_{2}}\right)\biggr],\nonumber\\
\emph{K}_{\perp}&=&2\chi\biggl[\frac{2\emph{k}}{E_{2}}-4\emph{k}\frac{\zeta}{E_{1}E_{2}}+ \emph{k}\zeta{\left(\frac{2}{E_{1}}-\frac{1}{E_{2}}\right)}\frac{1}{E_{2}}+2g_{\mathrm{s}}\frac{\zeta}{E^{2}_{1}}
\nonumber\\
&+&\emph{k}\zeta^{2}{\left(\frac{2}{E_{1}}-\frac{1}{E_{2}}\right)\left(\frac{1}{E_{2}}+\frac{2}{E_{1}}\right)}\frac{1}{2E_{2}}-g_{\mathrm{s}}\zeta^{2}\left(\frac{1}{2E^{2}_{1}E_{2}}-\frac{1}{2E_{1}E^{2}_{2}}+\frac{1}{2E^{3}_{2}}\right)\biggr].
\label{eq5}
\end{eqnarray}

In the above expressions,
$\chi=\langle r^{-3}\rangle_{3d}\emph{N}_{\mathrm{A}}\mu^{2}_{\mathrm{B}}$. Here $\langle r^{-3}\rangle_{3d}
(\approx 4.68$~a.u.~\cite{7}) is the expectation value of inverse
cube of the 3$\emph{d}$ radial wave function of $\text{Cu}^{2+}$ in
the  system studied. $\emph{ N}_{\mathrm{A}}$ is the Avogadro's number.
$\mu_{\mathrm{B}}$ is the Bohr magneton. Similarly, since the third- and fourth-order perturbation terms are neglected, i.e., only the first terms
in the right sides of equation~\eqref{eq5} are reserved, the above formulas are
reduced to those of the previous work~\cite{6}. Substituting these
values into equation~\eqref{eq5}, the Knight shifts are calculated and
collected in table~1. For comparisons, the theoretical results of
the previous work~\cite{6} based on the simple second-order $g$
factors and various adjustable energy separations are listed in
table~1.

\begin{table}[ht]
\caption{The Knight shifts and the $g$ factors for the
$\text{Cu}^{2+}(2)$ site in
$\text{YBa}_{2}\text{Cu}_{3}\text{O}_{7}$ of normal state.}
\vspace{1ex}
\begin{center}
   \begin{tabular}{cc cc cc cc}
   \hline   \hline
     & \ & $ g_{\parallel}$ & $g_{\perp}$ & $K_{\parallel}$ & $K_{\perp}$ \\
     \hline
   \hline
     & Cal.$^{a}$   & $2.345$    & $2.075$    & $1.282$    & $0.286$\\
     & Cal.$^{b}$   & $2.308$    & $2.057$    & $1.323$    & $0.242$\\
     & {Expt}.\cite{6,7} & $2.37(10)$ & $2.07(10)$ & $1.320(5)$ & $0.26(3)$\\
     \hline
   \hline
   \end{tabular}
   \end{center}
\small{
\begin{itemize}
\item[\textit{a}]Calculations based on the simple second-order $g$
factors and various adjustable energy separations in the previous
work~\cite{6}.
\item[\textit{b}] Calculations based on the uniform high order perturbation
formulas and the local structure of the tetragonal
$\text{Cu}^{2+}(2)$ site in
$\text{YBa}_{2}\text{Cu}_{3}\text{O}_{7}$ in this work.
\end{itemize}
}
 \end{table}

\section{Discussion}

From table~1, one can find that the theoretical Knight shifts
and $g$ factors of this work show reasonable agreement with the
experimental data, and these parameters are also suitably explained
in a uniform way.

In the present calculations, the relationships between the
Knight shifts (and the $g$ factors) and the local structure of the
tetragonal $\text{Cu}^{2+}$ site in
$\text{YBa}_{2}\text{Cu}_{3}\text{O}_{7}$
 are quantitatively established from the superposition model,
 and the shortcoming of the previous work~\cite{6} based on various
 adjustable energy separations is therefore overcome. Meanwhile,
 the high order perturbation formulas containing the third- and fourth-order
 perturbation contributions in this work are also superior to
 the simple second-order ones in the previous study.
 It is noted that a crude estimation of the contributions
 of the still higher (fifth) order perturbation terms yields $\sim \zeta^{4}/(E_{1}^{3}E_{2})$,
 in the order of $10^{-6}$ and being safely negligible. In addition,
 the previous results (Cal.$^{a}$) comparable with the observed values
 in view of the experimental uncertainties may be illustrated by
 the fact that the calculation errors happened to be canceled by
 the various adjustable energy separations~\cite{6}.

The positive anisotropy $\emph{K}_{\parallel}$ $-$
$\emph{K}_{\perp}$ is consistent with the positive anisotropy
$g_{\parallel}$ $-$ $g_{\perp}$. This can be ascribed to the
approximate linear relationships between the $g$ factors and the
Knight shifts. Further, the above anisotropies may be attributable
to the local tetragonal elongation of the $\text{Cu}^{2+}(2)$ site,
i.e., the relatively longer parallel $\text{Cu}$-$\text{O}$ bond as
compared with the four perpendicular ones. Thus, the ground
$^{2}B_{1{\mathrm{g}}}$ state and hence the anisotropic orbital angular
momentum and spin interaction between electron and nuclear can be
understood. The calculated anisotropy $\emph{K}_{\parallel}$ $-$
$\emph{K}_{\perp}$($\approx$ 1.081 in Cal.$^{b}$) of this work is
somewhat larger than that ($\approx$ 0.996 in  Cal.$^{a}$) of the
previous work~\cite{6}. This can be illustrated by the fact that the
present anisotropy $g_{\parallel}$ $-$ $g_{\perp}$($\approx$ 0.251
in Cal.$^{b}$) is smaller than the previous
result ($\approx$ 0.27 in Cal.$^{a}$)~\cite{6} and hence slightly
worse as compared with the observed value ($\approx$ 0.30 (20)~\cite{6,7}). The above discrepancies are ascribed to the
approximations of the theoretical model and formulas as well as to the
calculation errors in this work. In view of the experimental
uncertainties of the observed $\emph{K}_{\parallel}$ $-$
$\emph{K}_{\perp}$ and $g_{\parallel}$ $-$ $g_{\perp}$, the present
calculations can still be regarded as suitable.

\section*{Acknowledgements}
This work was supported by ``The Fundamental Research Funds for
the Central Universities'' under grant No.~ZYGX2010J047.

\newpage

\ukrainianpart

\title{Теоретичне вивчення зсувів  Найта $^{63}\text{Cu}$ \\ в основному стані
  $\text{YBa}_{2}\text{Cu}_{3}\text{O}_{7}$}

\author{М.К. Куанґ\refaddr{ad1}, С.-Й. Ву\refaddr{ad1,ad2}, З.-Г. Жанґ\refaddr{ad1}, Б.Т. Сонґ\refaddr{ad1}}

\addresses{\addr{ad1} Факультет прикладної фізики, Університет електроніки і технологій Китаю,
610054~Ченґду, Китай
\addr{ad2} Міжнародний центр фізики матеріалів, Китайська академія наук, 110016~Шеньянґ, Китай}

\makeukrtitle

\begin{abstract}
\tolerance=3000
Зсуви Найта  $^{63}\text{Cu}$ і  $g$ фактори в основному стані  $\text{YBa}_{2}\text{Cu}_{3}\text{O}_{7}$
в тетрагональній фазі вивчаються теоретично в єдиному підході з формул теорії збурень високого (четвертого) порядку
цих параметрів для $3d^{9}$ іона  в тетрагонально видовженому октаедрі.
Розрахунки кількісно корелюють з локальною структурою  $\text{Cu}^{2+}(2)$ вузла в $\text{YBa}_{2}\text{Cu}_{3}\text{O}_{7}$.
Теоретичні результати демонструють добре узгодження  зі спостережуваними значеннями, і покращення досягається шляхом використання
меншої кількості підгоночних параметрів ніж у попередніх роботах.  Знайдено, що значна анізотропія зсувів Найта є в основному атрибутом   анізотропії $g$
факторів завдяки взаємодії.
\keywords електронний парамагнетичний резонанс, ядерний магнітний резонанс, зсув Найта, $^{63}\text{Cu}$, $\text{YBa}_{2}\text{Cu}_{3}\text{O}_{7}$
\end{abstract}


\end{document}